\documentclass[11pt]{aastex}
\usepackage{emulateapj5}

\newenvironment{inlinefigure}{%
\def\@captype{figure}%
\noindent\begin{minipage}{0.999\linewidth}\begin{center}}
{\end{center}\end{minipage}\smallskip}

\slugcomment{{\sc Accepted by ApJ:} 1st April 2002}

\shortauthors{Steward Graduate Students}

\begin{document}

\title{On the Utter Irrelevance of LPL Graduate Students: An Unbiased
Survey by Steward Observatory Graduate Students}
\author{J.J. Charfman, J.B. Bsc, K.A. Eriksen, K. Knierman, A. Leistra, 
E. Mamajek, J. Monkiewicz, J. Moustakas, J. Murphy, \& J. Rigby}
\affil{Steward Observatory, University of Arizona, 933 N. Cherry Ave., 
	Tucson, AZ 85721, USA}
\email{}

\begin{abstract}
We present a new analysis of the irrelevance of Lunar and
Planetary Laboratory (LPL) graduate students at the University of
Arizona.  
Based on extensive Monte Carlo simulations we find that the actual number of 
useful results from LPL graduate students is $0\pm0.01~(5\sigma)$. Their
irrelevance quotient far surpasses that of string theorists.

\end{abstract}



\section{INTRODUCTION}
 
In a recent astro-ph submission, \citet{lpl02} claim that
the Lamentable Pathetic Lackey (hereafter LPL)
graduate students are dominant over Steward
Observatory graduate students. These findings are found to be
completely false, indeed laughable, in light of the evidence. 
LPL graduate students are found to be utterly irrelevant - more so than 
magnetic fields in most astrophysical settings.

The theoretical groundwork for LPL Graduate Student Irrelevance (LGSI)
was originally laid down as Fermat's Lamest Theorem (Fermat 1637; see
also Carrot-Top 1998 and references therein). The theorem was later included
by Einstein into a general theory of ``irrelativity'' (1918),
in which all beers taste the same independent of reference frame.
While much theoretical work has been done on the subject, only 
recently has empirical evidence been found to convincingly demonstrate LGSI. 

\section{Data}

\subsection{Sports}

 \citet{lpl02} claim LPL athletic superiority over Steward graduate
 students. They conveniently forgot to mention that {\bf two} events
 were to be scheduled: Volleyball and Ultimate Frisbee. LPL graduate
 students have consistently displayed shameless fear when asked to 
 compete in ultimate frisbee. Loss of bladder control was witnessed
 on at least one occasion (D. O'Brien, priv. comm.).

\subsection{Felonies}

As shown in Figure 1, LPL graduate students have a commanding
lead over Steward in terms of drug-related felonies. A Yakov-Smirnov
B-S test confirms this result at the 99.9997\% level. While this
evidence does not clearly demonstrate that Steward is cooler than LPL, it does
demonstrate that LPL grads are fairly slow, and easily caught.

\begin{inlinefigure}
\bigskip
\centerline{\rotatebox{270}{\includegraphics[width=0.6\linewidth]{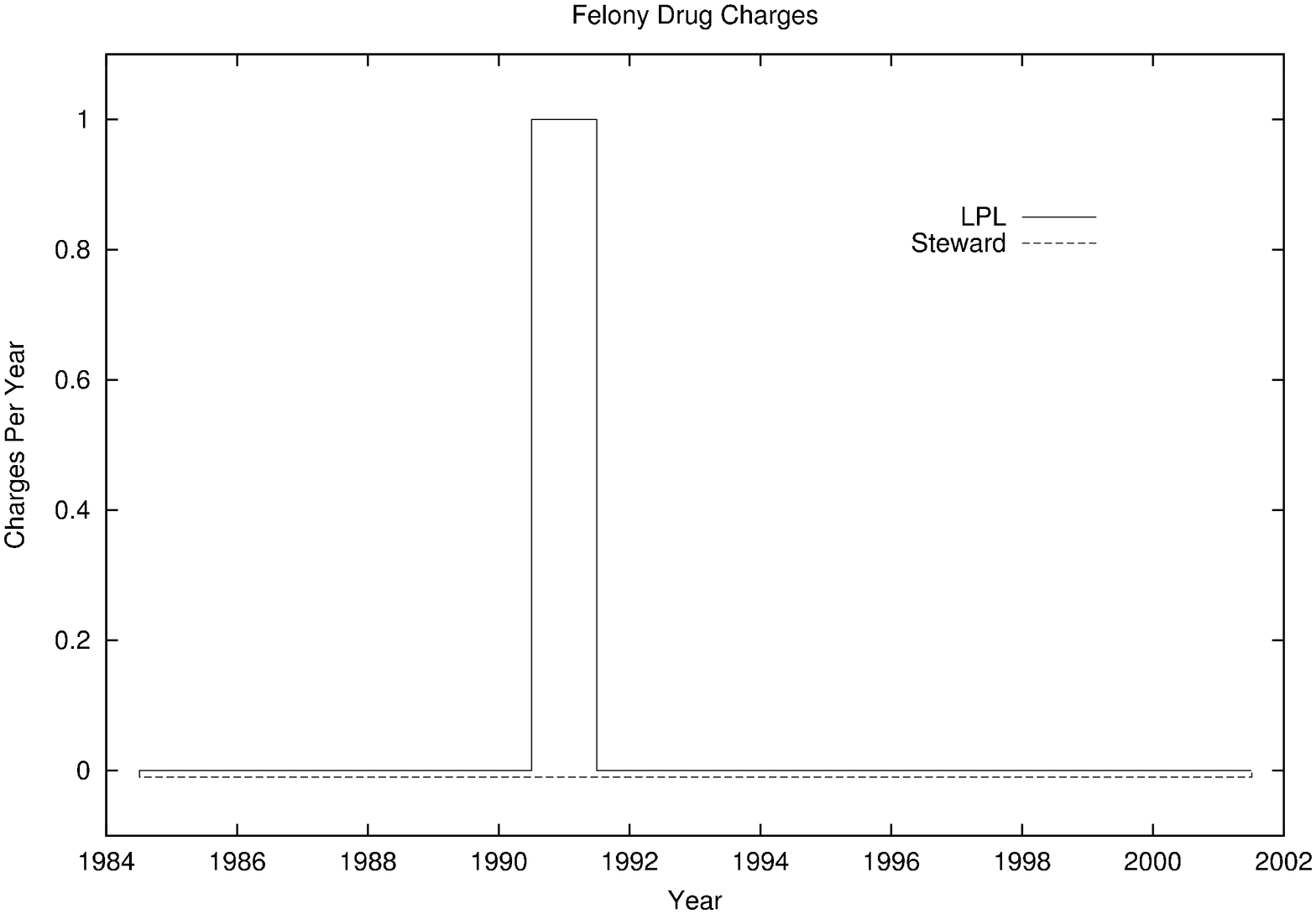}}}
\bigskip
\end{inlinefigure}

\vspace{-0.3in}

{\tiny\emph Fig.1~Felony drug charges versus time. This clearly demonstrates
that LPL grad students are slow, and easily caught.}

\subsection{Core Curriculum}

The academic core courses tackled by LPL graduate students have been
found to be farcical. 

\vspace{0.3in}
\centerline{LPL Core Curriculum}
\vspace{0.1cm}
\begin{tabular}{l|l}
\hline
501 & Introductory Rock Identification \\
502 & Intermediate Rock Identification \\
513 & Advanced Rock Identification \\
514 & Graduate Level Rock Identification \\
520 & Futurama Viewing\\
521 & Futurama Lab: Getting the Humor\\
542 & Astrobiology: The Chia Pet as an Unsustainable\\
    & Ecosystem\\
565 & Dodge-Ball$^a$ \& Paddy Cake\\
597B& Bratfest Greenhouse Gas Emission Laboratory\\
600 & ``Doctoral'' Nap-Time\\
\hline
\end{tabular}
(a) Due to popular demand a new Dodge-Ball-intensive\\
Planetary Sciences minor will be offered starting in 2004.

\subsection{Time Usage}

A pie-chart of LPL graduate student time usage is illustrated in
Figure 2. We define the number of useful papers from LPL grad students
$N$ using the famous ``Drake equation'': $N$ equals the number
of students $n_g$ divided by the volume of bad beer consumed $V_{Bud}$ times
the hours each student is awake $f_{zzz}$ 
(negligible to first order). We originally calculated
this to be unity, but later concluded that \citet{lpl02} did not merit
recognition as a useful paper.

Readers need not be reminded of Steward graduate student contributions 
such as the epic ``Super Huge Interferometric Telescope: A New Paradigm 
In Optical Interferometry'' \citep{rudnick99}, and the 
``The Effects of Moore's Law and Slacking on Large Computations'' 
\citep{gottbrath99}. 

\begin{inlinefigure}
\bigskip
\centerline{{\includegraphics[width=0.7\linewidth]{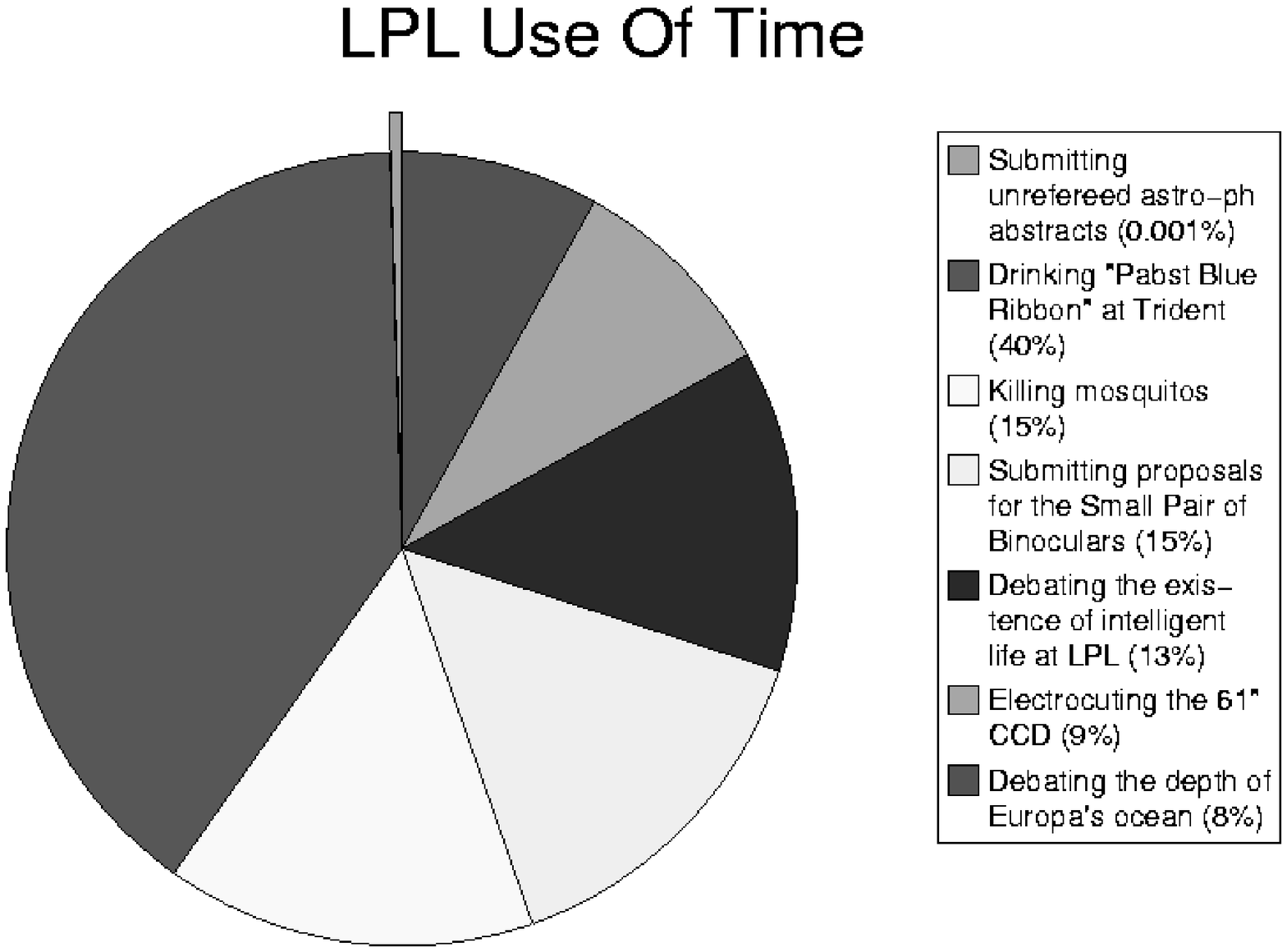}}}
\bigskip
\bigskip
\end{inlinefigure}

\vspace{-0.3in}
{\tiny\emph Fig.2~Mmmmmm.... pie.}

\subsection{Enrollment}

Only 2 students joined the LPL graduate program in fall 2001, 
compared to 10 at Steward Observatory. Only through begging and pleading 
on behalf of LPL bigwigs was the Lunar \& Planetary Lab building spared 
reallocation of office space to new Steward graduate students.

\subsection{Spelling}

\citet{lpl02} contained misspelled words such as ``dotoral'' and
``illigitimate''. This demonstrates the inability to run a standard
spell-checker. We later found that their paper {\it was} spell-checked, 
but with a Speak-And-Spell. A standard Speak-And-Spell contains 
80 vocabulary words and 14 barnyard animal sounds, so they were 
unable to check the spelling of their longer words. 
We can not rule out the possibility that the root of the
word ``dotoral'' is actually ``dolt'', however.

\section{Conclusion}
The utter irrelevance of LPL graduate students has been empirically
demonstrated.

\begin{inlinefigure}
\bigskip
\centerline{\includegraphics[width=1.0\linewidth]{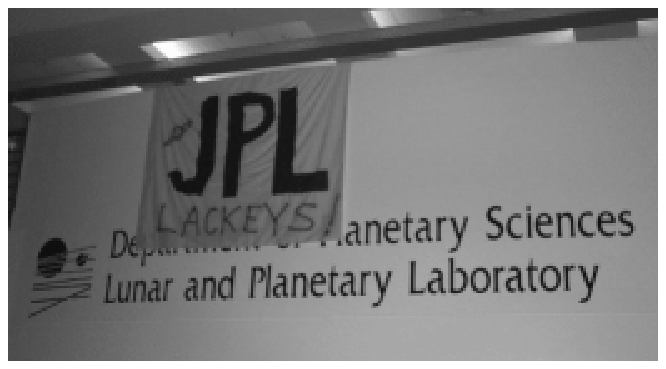}}
\bigskip
\end{inlinefigure}

\end{document}